**Title** OnRAMP for Regulating AI in Medical Products

*Author(s), and Corresponding Author(s)\* David C. Higgins\**

Dr. David C. Higgins
Berlin Institute of Health, Anna-Louisa-Karsch-Straße 2, 10178 Berlin
E-mail: dave@uiginn.com

Keywords: artificial intelligence, biostatistics, medical applications, methodology, regulatory

Medical Artificial Intelligence (AI) involves the application of machine learning algorithms to biomedical datasets in order to improve medical practices. Products incorporating medical AI require certification before deployment in most jurisdictions. To date, clear pathways for regulating medical AI are still under development. Below the level of formal pathways lies the actual practice of developing a medical AI solution. This Perspective proposes best practice guidelines for development compatible with the production of a regulatory package which, regardless of the formal regulatory path, will form a core component of a certification process. The approach is predicated on a statistical risk perspective, typical of medical device regulators, and a deep understanding of machine learning methodologies. These guidelines will allow all parties to communicate more clearly in the development of a common Good Machine Learning Practice (GMLP), and thus lead to the enhanced development of both medical AI products and regulations.

# 1. Introduction

Medical AI involves the application of machine learning algorithms to biomedical datasets in order to improve medical practices. Outside of a research context, medical AI must be built into products in order to deliver the desired improvements. Typical examples of medical AI products are: (i) the evaluation of one-or-more data modalities to support diagnosis, also known as clinical decision support (CDS), e.g. x-ray diagnostic systems; (ii) the statistical linkage of complex patient parameters to suggest optimal treatment paths, a riskier form of CDS, e.g. tumor biopsy analysis and treatment suggestion; (iii) the automated monitoring and detection or prediction of risk events, e.g. community-based monitoring via telemedicine of chemotherapy patients, or intensive care unit (ICU-)based monitoring for cardiac events; (iv) pre-/post-operative patient risk stratification for enhanced monitoring; (v) machine learning (ML)-based surgery navigation aids.(Basch et al. 2017; Ronen, Hayat, and Akalin 2019; Meyer et al. 2018; Kim, Li, and Kim 2020; Rajpurkar et al. 2017)

Any product with the potential to impact human health must be regulated before being placed on the market. In the USA, the US Food and Drug Administration (USFDA) is responsible for regulating medical products. In Europe, the regulatory framework is driven by the EU Commission and primary legislation, via a system of notified bodies. This means that in the EU medical AI products are regulated, depending on their intended use, as either *medical devices* (MD) or *in-vitro diagnostics* (IVD). The USFDA is currently the global leader in developing medical AI regulations. The initial focus of the USFDA has been on developing legal frameworks for regulation, running trials of multiple alternative regulatory pathways.(Health 2020) A recently updated USFDA position paper has declared a commitment henceforth to developing a customization of the existing, software-as-a-medical device (SaMD), regulatory framework for medical AI products.(Health 2021)

The academic community has begun proactively producing checklists and initial approaches for medical AI best practices, with a strong focus on reporting standards. The TRIPOD (Transparent Reporting of a multivariable prediction model for Individual Prognosis Or Diagnosis) Group is currently working on a checklist for reporting on clinical trials the data from which are subsequently used in the development of medical AI.(Collins et al. 2015; Collins and Moons 2019) MINIMAR (MINimum Information for Medical AI Reporting) represents a minimal list of fields, including AI descriptions, which must be reported in the development of a medical AI solution.(Hernandez-Boussard et al. 2020) The CONSORT - AI (Consolidated Standards of Reporting Trials - Artificial Intelligence) extension is an emerging standard for clinical trials evaluating efficacy of medical AI interventions, whereas its sister protocol SPIRIT - AI (Standard Protocol Items: Recommendations for Interventional Trials - Artificial Intelligence) provides AI-specific protocol guidance.(Rivera et al. 2020; Liu et al. 2020) Finally, Model Cards, and Fact Sheets are proposed as a means for better defining the acceptable uses of deployed AI systems.(Mitchell et al. 2019; Arnold et al. 2019)

Despite all of this progress, products incorporating medical AI are slow to appear on the market and typically fail to deliver on the promised levels of performance.(Benjamens, Dhunnoo, and Meskó 2020) Two key aspects, of this failure of translation, are: a lack of knowledge of regulatory practices, on the part of developers; and the absence of a best practices standard, required to produce safe and effective medical AI products.(Higgins and Madai 2020) The USFDA proposes Good Machine Learning Practice (GMLP) to describe a best practice standard, in clear analogy to the pharmaceutical drug development Good Manufacturing Practice (GMP). The USFDA position paper undertakes to support the development of such harmonized GMLP.(Health 2021)

This Perspective article proposes a first attempt at such a set of best practice guidelines for the regulatory compliant development of medical AI products. The focus is on linking the development roadmap to the regulatory dossier of documentation, which must be submitted as part of the regulatory approval process. A related (non-peer reviewed) article has been produced, including input from the current author, beginning with the regulatory audit and attempting to provide a templated approach particularly for the regulator.(Johner 2020) That article is expected to form the core of upcoming International Telecommunications Union and World Health Organization (ITU/WHO) best practice guidelines. Here the focus is on the perspective of the ML developer or data scientist, and how to structure the work systematically, in order to not miss vital steps which may later be considered necessary by the regulator.

## 2. Guidelines for Good Machine Learning Practice in Medical AI

In order to develop comprehensive best practice guidelines, for medical AI development, a basic introduction to common terms, and the structure of the regulatory process is first necessary. A glossary of useful technical and regulatory terms is presented directly following the **Conclusion**.

Three technical terms are of particular importance in medical AI development. Machine learning (ML) *algorithms* are computational procedures for learning from data. ML *models*, or trained ML models, are the output of the application of the ML algorithm to data. A trained ML model takes previously unseen data, similar to that in the training set, and makes predictions or classifications for it. Finally, ML *products* are products, which may be either physical devices or software, which use trained ML models in their operation. There are, additionally, a number of ML products that use ML algorithms whose outputs do not result in a trained model, e.g. clustering techniques. These will be discussed briefly in **Section 2.1.2** but otherwise rarely appear in medical AI products.

The purpose of the regulatory process is to ensure that medical products are both *safe* and *effective* for their intended use. Safety is evaluated from a risk perspective. The likelihood of misadventure is evaluated in terms of both proper and any potential improper use of the product. When the benefits of using the product sufficiently outweigh the risks of accident then the product can be certified for use. For medical products effectiveness is as important as safety. The product must demonstrably deliver any and all claims of medical benefits and must deliver a state-of-the-art level of performance when compared with competing solutions, including those only considered internally by the development company.

The process of evaluating the safety and efficacy claims of a new medical product follows a series of audits and inspections, with a particular emphasis on risk analysis and mitigation. As part of the application for certification the manufacturer must submit a regulatory dossier. This will include a technical file, or device master record, which describes the product and can be used to prove that it was produced according to the requirements of a quality management system (QMS). A design history file (DHF), which documents design decisions pertaining to the product, will also appear in the regulatory dossier. The inspector then examines the multiple steps of validation, assurance that the product meets the needs of all stakeholders, and verification, evaluation of the product compliance to specifications and requirements, which are described in the dossier.

Three regulatory terms, and the associated approaches to verification and validation, are of particular importance in medical product evaluation:

1. Intended use: The manufacturer's declaration as to all of the valid usages to which the product may be put. Ultimately the product as a whole must be evaluated, in a normative top-down manner, against its intended use.
2. Stakeholder requirements: A list of requirements in the form of statements on behalf of product stakeholders. These are derived, again in a top-down manner, starting with the intended use, and serve as an input to both verification and validation processes.
3. Technical *or* Product requirements: This is an exhaustive list of requirements, as to what a product should and should not do, which must be identified by the manufacturer developing the product, and serves as a checklist driven, bottom-up approach to verification.

While evaluation of both technical requirements and stakeholder requirements are largely driven by verification of conformity to the listed requirements, it is not possible to have a truly exhaustive checklist. Therefore, the auditor is primarily concerned with examining processes. Besides verifying that the technical requirements have been fulfilled, the auditor will examine whether robust processes have been established in order to formulate the requirements lists and ultimately to determine whether the intended use is likely to be safely delivered upon.

Following internationally agreed-upon standards of engineering is the easiest way to ensure that the needs of both verification and validation are satisfied and a product has been developed appropriately. The USFDA publishes a list of Recognized Consensus Standards for which a manufacturer may make a declaration of conformity in their premarket authorization filing. Applications based on non-recognized standards, or upon partial conformity to recognized standards, are also possible but require supporting documentation. The EU has historically maintained a list of Harmonised Standards, conformity to which was recognized as direct conformity to EU law. An updated list of standards, which conform to recent changes in EU regulations, has yet to be published.

The most commonly known standards are the International Organization for Standardization (ISO) standards. The International Electrotechnical Commission (IEC) provides some equally important, but less publicly known, standards. Further standards are provided by bodies such as National Electrical Manufacturers Association (NEMA), Institute of Electrical and Electronic Engineers (IEEE), and ASTM International. A brief summary of ISO and IEC standards most relevant to medical software development are presented in **Table 1**. A detailed description is beyond the scope of this article.

**Table 1.** Principal ISO and IEC standards of particular relevance for medical AI development.

| Principal standards for medical software | Standard title and relevant notes |
| --- | --- |
| IEC 62304 | Medical device software — Software life cycle processes. |
| IEC 62366-1 | Medical devices — Part 1: Application of usability engineering to medical devices.<br><br>Does not apply to clinical decision making. |
| ISO 14971 | Medical devices — Application of risk management to medical devices. |
| **Non Recognized Consensus standard.** A Declaration of Conformity (DOC) may *not* be submitted to USFDA based on this standard. | |
| ISO 13485<br>(Currently planned by USFDA for future inclusion as a Recognized Standard) | Medical devices — Quality management systems — Requirements for regulatory purposes.<br><br>This is a medical device specific version of ISO 9001. |

| **Non Harmonised Standard.** Cannot be used for EU certification. | |
|---|---|
| ISO/IEC 82304-1:2016 | Health software — Part 1: General requirements for product safety. |
| **Neither EU Harmonised nor US Recognized.** | |
| ISO/IEC 25010 | Systems and software engineering — Systems and software Quality Requirements and Evaluation (SQuaRE) — System and software quality models. |
| **Common standards which are not for medical software** These are *manufacturer* certifications, not product certifications. | |
| ISO 9001 | Quality management systems — Requirements. |
| ISO 27001 | Information security management. |

The following guidelines are divided into two sections. **Section 2.1** presents the bulk of the guidelines with a focus on product development topics. **Section 2.2** gives a much shorter, but highly necessary, overview of planning and preparing a clinical investigation for product validation.

## 2.1. Medical AI Product Development

In order to develop safe and effective medical AI products, which will pass the regulatory evaluation, the structure of AI development may be separated into six logically separate steps. Four of these apply to any medical AI product. Two of them are only relevant for specific, more complex, products.

1. Data curation: ML applied to healthcare is typically built upon labelled data. Data must have been properly acquired, securely stored, and the accuracy of labels is of vital importance. In rarer cases in which an unsupervised algorithm is used the trustworthiness of the data source is still an issue which must be quantified and guarded.
2. In-sample performance: A trained ML model must first be shown to perform well upon the original training data before testing on new data. The basic performance metrics, particularly on data which is similar to the training data, dictate the largest part of the risk-benefit analysis of deploying a new product. Beyond basic performance metrics these guidelines adopt two further approaches, namely algorithm-specific and dataset-specific perspectives to looking at in-sample data. The combination of all three approaches is what makes it safe, in the first instance, to deploy such a trained model in a clinical setting.
3. New data performance: In a real-world setting there is no guarantee that the data acquired for training perfectly mimics the deployment conditions. Therefore it is necessary to test the model's performance on newly acquired data. In general, this implies a clinical trial or study must be carried out. For anything but the lowest possible risk products, best practice further requires the development of an input plausibility model.
4. Output prioritization and resource planning: Similarly to how humans must learn to prioritize their actions, particularly in a clinical setting, complex medical AI products must prioritize their outputs. For example, a product which outputs a list of 100+ potential diagnoses, per patient, is likely to overwhelm the user. Therefore the outputs must be safely prioritized. Similar constraints apply to the process of sequencing operations when lab tests, or interventions, must be algorithmically scheduled. In order to do this safely it is necessary to develop risk models for the competing tasks and publish strategies which mitigate the risk-adjusted worst-case scenarios.

5. Adaptive AI: The USFDA has placed a priority on developing regulatory pathways for adaptive AI.(Health 2021) These are ML-based products which adapt to their deployment situation, updating their learned input-output mappings to better match the needs of the environment. As with any learning system there exists the possibility of unlearning. This must be strongly mitigated against or an adaptive AI system cannot be deployed.

6. Post-market planning: The manufacturer of medical products bears a legal responsibility for their product throughout the expected lifetime of the product. This means they cannot simply sell the product and then walk away. Rather they must plan and implement a strategy, for surveillance and updates, which incorporates the entire lifecycle of the product. For AI medical products issues such as shifting clinical standards and semantic drift are of particular consequence in this phase.

The sequence of these steps follows the natural flow of a data or ML project. Data curation, despite being foundational, is largely a software engineering task so the focus, in this guide, is on the data-specific aspects. In-sample performance forms the core of a typical data or ML project and similarly forms the bulk of the guidelines. New data performance and post-market planning are regulatory requirements which, while not necessarily technically demanding, deserve separate consideration. The sections on output prioritization and adaptive AI are only necessary for specific types of products and, otherwise, may be skipped.

Importantly here, the focus of these best practice guidelines is on certification which follows from documentation. In a structured product development path it is common to conduct exploratory analyses, comparing many different modelling approaches, on early data

sets.(Higgins and Madai 2020) The level of documentation required for product certification is extremely high. Therefore, a balance must be struck between fully documenting paths, which are unlikely to subsequently be pursued, and omitting documentation of early attempts. The former will lead to deeper trust from the side of the auditor, but the latter is completely reasonable until the point at which an intended use is defined.

The particular form the documentation takes is outside the scope of these guidelines. It typically involves documented standard-operating-procedures (SOPs), for all product development processes, accompanied by evidence that the SOPs were actually followed. Any decisions made, or considered, should be documented from the perspective of: paths considered, reasons for choices, potential risks, and risk mitigation of the chosen paths.

A summary of the sections from these guidelines, and their respective subtasks, is presented in **Table 2**.

**Table 2.** Summary guide to build and evaluate the ML components of a medical AI product.

| Macro-task | Sub-task | Details |
|---|---|---|
| Data curation | Data storage | Should be done using appropriately certified software and hardware systems. |
| | Data acquisition | Evaluate digitalization approach for accuracy and plausibility. Evaluate data population for relationship to intended use population. Evaluate data for 'quality' relative to target operational deployment settings. |
| | Data labelling | Should be done via medical quality certified software approach. Evaluate expert-labeller performance. Consider multi-fold approach. Justify conflict resolution approach. |
| In-sample performance | Basic model performance reporting | Sufficiency of randomization justification. Evidence for avoidance of overfitting. Intended use: automation vs exploration. Appropriate performance measures. Cross-validation performance. |
| | Algorithm properties perspective | Justify the ML model used. Sufficiency of the data set. Subpopulation analysis. Sufficiency of the data at clinically relevant decision thresholds (see **Figure 1**). Evaluation of clinical relevance of decision criteria. Evaluation of appropriateness of hyperparameters. Sensitivity analysis of performance to training set size. In-distribution performance evaluation (e.g. via synthetic data). |
| | Data properties perspective | Handling missing input values and data imputation justification (see **Table 3**). Handling missing outcome values. Handling of discontinuous input data. Input representation transformation justification. Feature engineering justification. Is convergence assumption realistic for non-Gaussian data? Safe failure modes. |

**Table 2(-cont'd).**

| New data performance | Detection of out-of-distribution input. | Input plausibility model. |
|---|---|---|
| | Safe handling of out-of-distribution input. | Error handling routines. |
| | Clinical trial. | Model calibration. Study population description matches intended use. Study population shows an unbiased cross-section of society. Trial endpoints match intended use. Use of an appropriate control. ML objective follows intended use. UI/UX evaluation. Reporting follows a standardized checklist approach such as CONSORT-AI or SPIRIT-AI.(Liu et al. 2020; Rivera et al. 2020) |
| Output prioritization and resource planning | Blending of outputs. | How is the blend achieved? i.e. What is the ranking method, or relative weighting for different output categories. Evaluation of relative performance for different outputs. Risk evaluation for over-/under-diagnosis and the severity of the associated outcomes. |
| | Communication of blended outputs with users. | How is the primary output communicated relative to communication of secondary outputs? Risk evaluation that users will misunderstand the relative performance reliability for different outputs. |
| | Resource planning. | Task identification performance. Task prioritisation performance. Overall evaluation in terms of intended use. |

**Table 2(-cont'd).**

| Adaptive AI | Risk analysis | Risk-benefit analysis must specifically justify the use of an adaptive algorithm. Quality assurance of any teaching or learning signal. Detailed algorithmic analysis for potential to introduce systemic biases. |
|---|---|---|
| | Safe bounds on learning performance. | A minimal quality must be guaranteed. |
| | Appropriate performance monitoring modules. | Existence of a non-adaptive performance monitor. Robustness analysis of the non-adaptive performance monitor. |
| Post-Market planning | Surveillance. | Paper based methods. Telemetry streaming allowed? |
| | Software updates. | Minor vs major updates. Deployment methodology. Support for old versions until removed from use. |
| | Shifting clinical standards / Non-stationarity of data inputs over time. | Semantic drift. New diagnoses. New input fields. |

*2.1.1. Data Curation*

Data curation involves the acquisition, labeling and storage of reliable, unbiased, and high-quality data. An algorithm which is not built upon a strong basis of data curation is not fit for purpose. The data curation processes must not only be well planned and executed, they must also be documented with particular focus on risks.

Data gathering and storage processes must be robust to potential error-introducing steps. In particular, data storage devices, and software, must themselves follow basic risk mitigation norms. Such systems, even sourced from third-party vendors, must be backed by a Quality Management System (QMS) or separate quality assurance (QA) certification.

In an attempt to acquire data for medical AI it is common to transcribe paper-based medical records onto digital systems. This is an error-prone process and should be evaluated for the likelihood that it introduces errors into the data. For example, in drug development it is common to have a minimum of 2 individuals enter every case report separately, into digital form, and then an analysis is performed for accuracy before a single coherent data set is 'locked'.(Pitman 2019) Where clinical decisions are a factor, these are frequently reviewed by multi-disciplinary panels before incorporation into the training set.

Data which does not reflect the intended use is a form of bias. Two mistakes occur commonly here. First, it is usual to kick-off projects at world-class centres of excellence; such centres rarely see a 'normal' cohort of patients. If the final product is to be deployed in typical patient care, then data must be gathered from precisely such a population of patients. For example, a predictive model trained on a high-risk patient population cannot be certified to safely make predictions on low-risk patients without considerable further data acquisition and testing. Another common mistake is in gathering 'too perfect' a data set. There is considerable scientific value in having a high quality data set. However, unless statistics are gathered as to which patterns of entries are missing from the intended deployment setting then this will also lead to a biased data set.(Hand 2020) For example, a particular biomarker test might be included in the 'perfect' data set but is not standard-of-care in typical clinical practice. The developed model may be overly dependent on this biomarker leading to biased, and likely inferior, performance in real-world usage. It is not possible to deploy an uncorrected model, which has been trained on biased data, on any other patient group.

Finally, medical AI is usually developed on labelled data. Therefore it is vital to ensure that the labelling process is correctly handled. This contains both a technical and a human aspect. As with storage, the software for labelling should be certified to at least a medical software quality standard. Similar to the digitization process, it is common here to use multiple experts to independently label each data entry. In this case, a thorough analysis must be presented as to the degree of consensus among experts and how cases of disagreement are handled.

*2.1.2. In-Sample Performance*

How well does the trained model work on the data gathered so far?

Performance of a trained model on the data set from which it was derived is referred to as *in-sample performance*. There is additionally a second, statistical meaning of in-sample which refers not to the sample data itself but to the data generating process from which the data is understood to have derived. In order to disambiguate the two meanings, this latter meaning will be referred to as *in-distribution performance*.

Best practice for the development of any machine learning (ML) model is the use of a segregated, randomly selected, test set, and the use of cross-validation for any parameter tuning on the remaining training set.(Russell and Norvig 2016) Randomization of medical data sets carries a number of difficulties, notably due to different patient cohorts and inconsistencies in medical record keeping. It is, for example, incredibly common for the same patient to appear multiple times in a data set with different patient IDs. This means particular care must be paid to the quality of the randomization, preventing leakage of information between training and test sets, while preserving the desired predictive modalities across the sets. In reporting the development of the AI product, the sufficiency of the train-test or cross-validation set randomization must be justified.

Since modern ML models may have an internal dimensionality high enough to perfectly capture and fit every data point on which they trained, the use of a test set is extremely important to examine generalization and prevent overfitting.(Zhang et al. 2017) If the test set has been correctly constructed the performance results on the training set, e.g. via multi-fold cross-validation, should form a narrow distribution around the test-set metric. Some medical data sets are too small to reliably allow such an approach, in which case considerable justification must be made about risk-benefit payoff, and a lack of alternative methods, before a product built on such data may be considered for certification.

Software tools which incorporate machine learning have many purposes. An important distinction however is in whether the ML is to be used for the purpose of automation or for exploration. Typically medical products are not certified for discovery-style exploration. The risk of misuse is too high to justify certification. Therefore, although ML is very good at this task it remains the domain of scientific research. Any software used in such a mode is then at the legal liability of the user and is no longer the responsibility of the developers. An exception is a medical device whose intended use is scientific, rather than directly medical, and which will be operated by a user with sufficient scientific training to be able to interpret the output correctly. For example, a clustering technique, the output of which, at the discretion of the analyst, may subsequently contribute to a histopathology report. While scientific tools have historically been exempt from regulation, recent changes in EU law include them under medical device regulations once any competing certified product appears on the market.

Appropriate model performance metrics are algorithm and application specific. What is important is that they must be thoroughly pursued, their particular use is justified, and all of their values reported. For example, for classifier systems selective reporting of sensitivity and specificity without reference to positive predictive value (PPV) and negative predictive value (NPV) is inappropriate. Rather, it is desirable that the confusion matrix and all derived measures are to be reported. As a second step, it is then perfectly understandable for the regulator that sensitivity and specificity are of particular interest, from a risk analysis perspective, for screening purposes whereas PPV and NPV are preferred for individual diagnostic or treatment decisions. Similarly, metrics such as accuracy and area-under-the-curve (AUC) may be reported, but they are unlikely to be sufficiently informative to evaluate most products under their intended use evaluation. The reason for this is due to the clear asymmetry of risk, between false positives and false negatives, in particular medical contexts.

Full reporting of model performance during any cross-validation performed, e.g during hyperparameter tuning, should be the norm. For small data sets, in products presenting a low risk profile, it may be allowable to report average model performance rather than test-set performance. Correct model comparison methods, for MxN-fold cross-validation, are particularly tricky. Correct model comparison equations are presented in papers by Nadeau and Bouckaert.(Nadeau and Bengio 2000; Bouckaert and Frank 2004)

The first step to reporting on in-sample performance is the establishment of basic machine learning norms. This must, at all points, be thorough, detailed, and must openly compare with state-of-the-art performance. Beyond this there are two further technical issues on in-sample performance, which only the machine learning team have the expertise to address. They are, however, frequently overlooked in consumer-tech approaches to ML development. These issues are (i) the technical implications of the chosen ML algorithms, and (ii) the statistical properties of the underlying data sets. These issues will be addressed in the following two subsections.

*Technical Implications of the ML Algorithm:*

Considerable theoretical work has been carried out into the algorithms behind machine learning techniques. However, an understanding of these theoretical foundations is much less widely available. By evaluating the product development through the perspective of the algorithm enormous insights can be gained into the evaluation of in-sample performance.

A thorough description of all ML algorithms applied to the data, their relative merits, and why the final algorithm was chosen is a prerequisite for any regulatory evaluation of a ML product. The chosen algorithm must be justified as being the most appropriate for the application. The justification cannot be based solely on the preferences and available skills in the development company. Further, alternative means, i.e. non-ML methods, must also be considered and if any are considered superior to the ML-based approach, from a risk-benefit point of view, then the ML-based product cannot be certified.

Within a particular ML model class it is often possible to evaluate whether the data set is sufficiently large to support the reported model performance metrics. Although it is never recommended to estimate *a-priori* how large a data set will be required for a particular machine learning task, it is often possible to evaluate algorithm performance *post-hoc* using a combination of heuristic-based approaches and actual theoretical results to decide whether the estimated performance metrics are accurate or not.(Hua et al. 2005)

The heuristic approach, to evaluate the general applicability of the model performance metrics, is essentially a modern data science validation approach. Particular focus should be on identifying whether there are specific sub-groups in the training data for whom the ML model works better. This must be combined with a risk evaluation. It is frequently *less risky* to have a higher estimated variance on patients in extremal medical categories (e.g. completely well or almost certainly dying), since their outcomes are somewhat pre-destined. It is much riskier to have higher variance for predictions regarding patients whose data lies close to a clinically relevant decision threshold. This is typically a problem of data acquisition in the original data samples. What is frequently observed in real-world clinical applications is training data which has very high sample size, and low variance, for the extremal data points and very low sample size, and high variance, in the intermediate zone (**Figure 1**). This leads to easy automation of *easy* tasks, but incredible difficulties in building a good model close to the clinically relevant decision boundaries. This problem should be identified early by the data team and a strategy, such as targeted data acquisition or some form of hybrid machine learning model perhaps containing an explicit model for uncertainty, should be developed to mitigate this phenomenon.(Sensoy et al. 2020)

**Figure 1.** Cartoon illustration based on experience of real-world medical AI projects. Availability of data (red; left axis) is frequently inversely proportional to the target variance (blue; left axis), or task difficulty, in biological prediction tasks. Meanwhile the greatest possibility to influence patient outcomes (green; right axis), i.e. the intermediate-category containing the most clinically relevant decision threshold, lies precisely in the zone of maximal variance and lowest data availability.

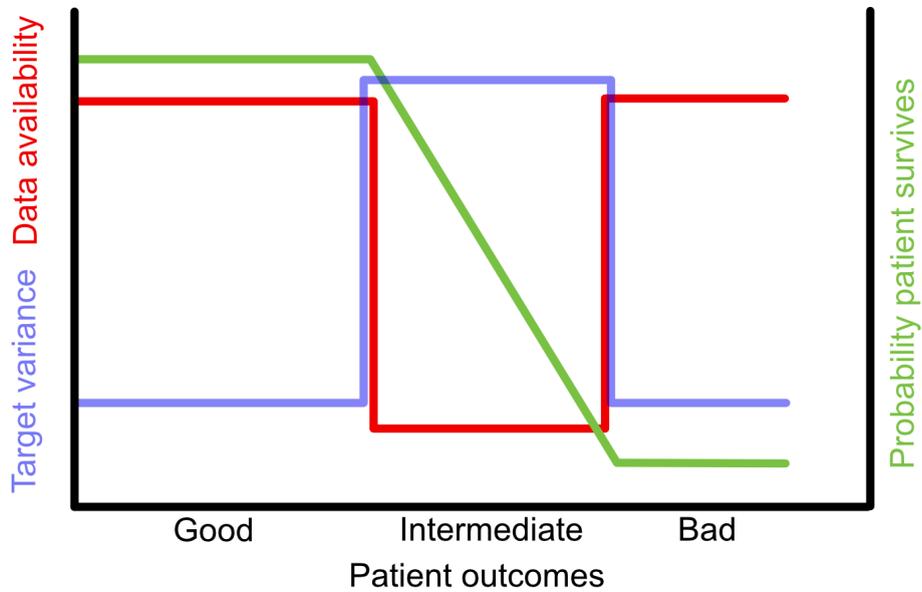

A more theory-based approach, to evaluating the model performance metrics, begins from the characteristics of the ML algorithm. Every algorithm has specific characteristics which are core to the analysis of expected real-world performance of the developed tool. For example, knowledge that tree algorithms split on a single feature at a time should lead to an investigation of the clinical relevance of the tree nodes. Shapley values are particularly of interest here.(Lundberg et al. 2018) Their use, however, must be balanced by an awareness that while post-hoc explanations of model decision-making criteria are a popular research topic their accuracy is highly debated.(Merrick and Taly 2020; Rudin 2019) Similarly, when hyperparameter tuning has been performed the appropriateness of the chosen hyperparameters must be justified and alternatives considered should be reported. Finally, a numerical sensitivity analysis of model performance relative to training dataset size should be reported.

The final application of specific algorithm knowledge is in evaluating how the trained model is likely to perform on 'similar' data. This follows the definition of *in-distribution* performance (see beginning **Section 2.1.2**). A training and test set are always finite samples even if they form excellent approximations of the underlying distributions. This means there may be many points of interest, which might be generated by the same data generating process and which might lead to as yet undiscovered results from the trained algorithm. A good evaluation of the algorithm will attempt to answer the question of how the trained model is likely to perform on data which is generated from the same data generating process but which is not identical to already tested points. The ideal evidence for in-distribution performance is whether the solution space can be shown to be some kind of smoothly differentiable manifold. Theoretical results currently fall short of being able to provide such assurances, however a numerical sensitivity analysis using synthetic data can give useful indications.

In general, an evaluation of in-sample performance under the lens of the properties of the algorithm will lead to: discovery of numerous problem cases which must be mitigated; holes in data gathering which must be filled; and, insights into how the trained model is likely to work on similar patient cohorts. The absence of a history of such insights and their remedies, in the formal documentation, is a strong indicator of insufficient quality control processes or adherence to said processes.

*Statistical Properties of the Data:*

Machine learning algorithms converge differently depending upon the structure of the underlying data distribution. Biomedical data sets frequently contain missing input values, missing outcome information, discontinuities in the encoded data, and statistically significant patterns which do not translate well to real-world usage. At a deeper level, not all data follows well-understood distributions, such as the normal distribution. All of which leads to serious model, and product, design implications which must be addressed when building a medical AI solution.

Missing input values are a common feature of medical data. In medical imaging this is less of a problem, and can perhaps be remedied by either value imputation or an insistence on retaking an intact image. However, a typical medical history is frequently sparse, and driven by clinical imperatives at the time of presentation. The presence or absence of a lab result may be more indicative of the final diagnosis than the actual lab values reported. A well constructed medical AI system needs to be able to deal appropriately with these missing values. See **Table 3** for some examples.

**Table 3.** Appropriate data imputation in biomedical applications can depend on the intended use. Sometimes it is dangerous to impute missing values (eg. weight change during pregnancy). In other cases, the form of imputation used can directly interact with the form of the algorithm.

| Data Imputation | Example |
| --- | --- |
| Some parameters can be imputed from previous visits. | Age, height, weight |
| The same parameters cannot be imputed under specific clinical conditions. | Weight will change during pregnancy and the change may be clinically relevant. |
| Some parameter values should be switched to a binary flag (e.g. present / not present). | Was a lab result above a proven clinically relevant threshold? |
| When marking missing data with a 'NA' code, this code should not adversely interact with the algorithm. | If the actual values range from 0-5 then using '-1' for missing values will bias regression-based algorithms. |

Unfortunately, obtaining outcome data from medical results can be both difficult and costly. In certain cases, e.g. pregnancy, it is reasonable to assume that a complete medical record set will be gathered for most patients. But in most illnesses, even in cases where the patient subsequently returns to the same doctor or clinic, it is rarely possible to ascertain whether the original diagnosis-treatment combination was successful or not. This is important from a labeling point-of-view. It is only possible to build products which automate the diagnostic or treatment decisions of the clinician if the outcomes are known. The next best alternative is that an expert committee must decide, based on the data, what diagnoses or treatments should be assigned. In either case, an analysis of the training data must be performed with particular focus on evaluating the risk that the data points with well-labelled outcomes form a non-representative subset of the general population, which would lead to a biased real-world performance.

Discontinuities appear frequently in biological data. This may be the result of the monitoring equipment, which is designed to work on timescales relevant to steady-state values e.g. early blood-glucose monitoring devices. Or, discontinuities may represent biological on-off processes which switch the production of certain hormones, and other biochemicals, either entirely-on or entirely-off in the body. In general, machine learning algorithms operate under assumptions of continuity which leads to inconsistent outcomes when fitting discontinuities. The appropriate use of a kernel smoother function transform in the feature space representation can greatly aid in combining discontinuous data.(Hofmann, Schölkopf, and Smola 2008)

Frequently, the easiest way to improve the performance of a machine learning algorithm on biomedical data is the use of input representation transformations. Simple transformations include normalization, transforming into another set of dimensions, and the use of input encoding values. Normalization supports the ML algorithm by reducing the numerical computing difficulties of handling inputs with vastly different value ranges. Whereas dimensional transformations and the use of input encoding values provide support respectively, by leading to easier separation of data to be learned, and by building-in knowledge, such as previous diagnoses, which must otherwise be learned from base data. Each of these methods, when used carefully, represents only a low risk to the final product. However, the use of more complex, particularly manually generated, engineered features must be very carefully analysed for risk.

The problem with complex feature engineering is best motivated by an analogical example. There is a well-known, but perhaps apocryphal, story of a product for AI-driven medical radiography which showed excellent in-sample performance but further investigation revealed that it was relying on spurious information which spanned both the training and test sets. The story relates that the spurious information was a unique tag, in the borders of the images, which showed that the image originated from a clinic which specialized in a particularly rare clinical condition. The algorithm was able to leverage this statistical association to appear performant at diagnosing that rare disease. In the ML literature such a system is referred to as a 'Clever Hans'.(Lapuschkin et al. 2019) The problem with complex engineered features is that they curate such hidden associations, and they do so in a manner which is considerably more difficult to trace than that in the example. Therefore, feature engineering must be motivated by either numerical computing factors or by demonstrable biological association between the engineered feature and the target outputs.

Much of statistical theory is based on the normal distribution of data. It is possible to evaluate approximately how many data points are needed before reasonable estimates of the mean and variance of such distributions are obtained. In the real-world, biological data is frequently non-normally distributed.(Limpert, Stahel, and Abbt 2001) From a statistical analysis perspective, it has been shown that if it is not possible to distinguish between a log-normal distribution and a normal distribution then the statistical properties of the log-normal distribution must be assumed.(Taleb 2020) In such a case, the convergence properties for statistical methods are slowed to the point where no reasonable amount of data can assure convergence. ML algorithms have different convergence properties from statistical algorithms. But still, they can be expected to insufficiently converge for fat-tailed distributions. Recent work has called this incomplete convergence a form of model underspecification.(D'Amour et al. 2020) The risk is that the variance will always be underestimated and as a consequence so will the error. If the algorithm cannot reasonably converge then it is impossible to validate it for clinical deployment.

The best advice here is to use more complex, but also more robust, heuristics in evaluating how much data to employ and, more importantly, to fall-back on product designs which avoid the catastrophic effects of massive failure. This point, about failure modes, is extremely relevant for in-sample performance evaluation. Ultimately, every tool fails at some point, therefore the product should be designed to fail in a graceful manner. This topic will have even more relevance in the following sections.

*2.1.3. New Data Performance*

Assuming that a machine learning model, for medical AI, has been successfully built and has been shown to work reliably and with sufficient efficacy on a large data set. The next task is to validate how the model will perform on other data sets. Here the risk is that despite excellent model performance under the original data generating process, this will not transfer to real-world data generation situations.

Essentially, there are two complementary approaches here. First, a clinical trial needs to be conducted in order to demonstrate that the performance statistics reported for in-sample performance are reproduced in a clinical situation.(Higgins and Madai 2020) This is also an opportunity to evaluate model calibration, ie. how well the model performs across all cohorts, the definition and calculation of which is not trivial for a predictive model.(Archer et al. 2020; Riley et al. 2020) The trial itself will follow normal clinical trial protocols, such as definition of trial participant selection criteria and intended use, a discussion of which is in **Section 2.2**. Reporting of the trial should follow standardized checklist-based approaches, such as CONSORT - AI and SPIRIT - AI.(Liu et al. 2020; Rivera et al. 2020)

The task of the development team is to determine bounds for what the valid population, on which the product will be used, looks like. This can, of course, drive the clinical trial design process. But, more importantly, this will lead to the development of an input plausibility model. This is the second of the two complementary approaches.

Data which is submitted to the trained model must first be evaluated for plausibility. This will ensure that data entry errors are caught, and rectified, before any model prediction is run. It will also, potentially, catch patients whose clinical condition does not match the conditions under which the model can be reasonably expected to operate.

The safety issue posed, by inappropriate application of a ML model, cannot be overestimated. A machine learning model should not be deployed on a population which is different from the one on which it was trained. Despite the term AI incorporating the word intelligence, machine learning cannot self-correct to cope with even minor situational differences for which it has not been trained.

A simple input plausibility model will set maximum and minimum values for each input field. A more advanced model will use a support-vector-machine (or similar) which has been designed for the task of one-sided classification.(D. Tax 2001; D. M. J. Tax and Duin 2001; Stephan 2001) That is, a model which has been trained to recognize data similar to the original training data and may, as a result, be used to detect anomalies. The most advanced techniques, even if they do not explicitly use a Bayesian method, will reflect the Bayesian question: What is the likelihood that the new data point is from the same data distribution as the original training data?

The absence of an input plausibility model must be motivated by a thorough risk evaluation of the lack of potential for patient harm. In contrast, the presence of a sufficiently advanced input plausibility model may be taken as evidence that the product can safely operate even for intended uses in which the potential for patient harm would otherwise be significant.

*2.1.4. Output Prioritization and Resource Planning*

As more and more complex medical AI systems are developed, the models involved will develop towards increasing levels and hierarchies of models and automation. A single task medical AI model can be evaluated by relying only on the sections leading up to this one. In this section, the consideration is on the situation where the overall product might be used for more than one medical condition, or application, or across multiple medical contexts. Such products might someday range from electronic health record integrated clinical decision support tools, which present a ranked list of potential diagnoses for a patient, all the way to automated artificial general intelligences (AGI) which take symptoms as input and directly prescribe treatments without the intervention of a human operator.

In the simplest example, a single trained ML model takes the data from a patient and evaluates the data for the likelihood that it represents a single clinical indication. A second ML model is trained for a second indication. And this process of training separate models continues for each indication which is considered of interest by the product developers.

The difficulty, in this example, is that in a product the output of each of the individually developed models must be aggregated and presented to a user, such as a clinician, in a coherent manner. In order to do this, a ranking or prioritization algorithm must be developed. A naive approach might involve ranking the indications based on how certain each model is that its diagnosis is correct. To do this correctly, the individual models must have been trained to output such a likelihood and not just a binary classification.

This approach still leads to two difficulties. First, since different indications have very different prevalences the accuracy of the different models is not directly comparable. And second, different indications represent very different clinical risks from a medical point-of-view. That is, for some medical conditions it is important to (over-)respond early rather than rely on the fact that such a diagnosis is rare, but treated too late will be fatal. This is a core concept in the method of differential diagnosis in medicine.

Moving beyond this simple example, of hybridising the outputs of separate models, the same issues emerge when directly validating multi-class classifiers and Bayesian Network approaches. These techniques overcome some of the technical issues of comparing performance across multiple outputs. But such approaches still suffer from the prioritization issue. These are issues which must be considered in terms of the intended use, and how this use is communicated, first via the device sales communication process and then through the user interface, to the end users. User testing must be carried out, and reported upon, in order to evaluate the risks for miscommunication and subsequent adverse outcomes.

Finally, when patients present to a doctor the clinician must quickly evaluate the entire patient presentation. This incorporates both a high- and a low-level interpretation of symptoms and a priority based approach to investigation and treatment. At some point 'smart' algorithms, which similarly decide where medical resources are to be concentrated, will have to be evaluated. For such an algorithm to work appropriately, it must operate a paradigm which evaluates the ongoing task environment and ranks which sub-algorithms, or trained models, are most important to subsequently proceed with. Q-learning is an example of a commonly known technique used in this area today.(Watkins 1989) In computational neuroscience and research into AGI such an algorithm embodies a task identification and/or switching process. The development of such algorithms should attempt to follow basic theoretically advantageous properties, such as the probably-approximately-correct (PAC) framework, but where the target or goal is fuzzier than that of a stand-alone tool.(L. G. Valiant 1984; L. Valiant 2014) The evaluation of this top-level process (frequently referred to as a module) must also be carried out, and will follow similar procedures to those referred to for the individual machine learning models. From a legal perspective, such an autonomous system is regulated according to its ability to safely and efficaciously serve its intended use.

*2.1.5. Adaptive AI*

A recent USFDA discussion document has made it clear that regulatory bodies are not just considering static, or 'locked', ML-based products, with a focus on when they can be changed without undergoing re-certification, but are also willing to engage with developers planning adaptive AI solutions.(Health 2020) In this case, the product is assumed to have an initial base-level of performance but is capable of modifying its performance characteristics based on the pattern of inputs received over time. This is probably best understood as a similar technology to that present in smartphone predictive text keyboards, which can adapt to the user's word-usage preferences, performing better over time.

From a machine learning point-of-view, it is important to note that the meaning of adaptive here is not referring to techniques with hidden internal states, such as Hidden Markov Models, recurrent Neural Networks (rNNs), etc., where the internal state for an ongoing prediction is updated based on newly inputted data. For example, this is *not* a real-time ML model where the prediction for a patient may change immediately based on their answer to a symptomatic question. Such models, despite their technical complexities, are regulated in exactly the same manner as all other 'locked' ML models.

Rather, the USFDA definition of adaptive AI should be read to mean models which are locally updated without interference from the developers. That is, methods which autonomously retrain the model based on patterns of data inputted over time. Common techniques for adaptation include reward learning and Bayesian methods. Both of which are, of course, compatible with the development of 'locked' ML-models but are also utilisable for ongoing performance improvements. For example, a reward learning method could be used in combination with accurate lab-based patient outcome data to advance the point in time, at which a particular illness may be accurately detected, to an earlier stage in the patient journey. Alternatively, a Bayesian approach could modify its internal expectations of 'normal' patients based on experience of local conditions. That is, each clinic would have a ML model which over time would adapt to issues such as regional incidence rates of particular illnesses or comorbidities.

Clearly, in the reward learning example, the strongest issue is one of validity of the teaching or feedback signal. This issue is the same as the labelling quality control issue posed in **Section 2.1.1**. The product manufacturer is responsible for ensuring the quality control of the feedback signal. The USFDA has declared that they will take a risk-based approach to this quality control issue. That is, if the risk of causing harm is sufficiently low, relative to the expected patient benefits, then such an approach may be granted approval.

The Bayesian example contains an important demonstration of the hidden potential for harm posed by adaptive AI products. Adaptation to local conditions contains a strong risk of algorithmically incorporating systemic bias into the model. That is, rather than correctly detecting a higher local incidence of a disease, which only affects subgroups of the national population, the product may falsely try to 'normalise' the incidence rate to national levels which are below those experienced locally.(Obermeyer et al. 2019) Therefore the adaptive component of the AI must be evaluated against the risk of increasing rather than decreasing already existing socio-economic and genetic biases which exist in the population.

With adaptive AI the most important regulatory aspect is in assuring that a minimum level of performance is always guaranteed. In any coupled learning process it is always possible to learn how to do things badly. The product should therefore enforce two safeguards: (i) bounds on the magnitude of the potential learning process; and (ii) strengthened monitoring processes. The goal here is to certify the product as always performing above some minimal acceptable level, while offering the possibility of improved performances.

*2.1.6. Post-Market Plans*

As part of the regulatory package a developer of a medical AI product should also expect to submit plans for post-market operations. Two aspects of these plans are relatively clear and are similar for any medical software device: surveillance and dealing with software updates.

Surveillance means the post-market monitoring for any adverse events which occur through the normal use, or misuse, of the product. This is separate from crash monitoring, typical in non-medical device software. Indeed, in most jurisdictions paper-based inbound reporting must still be supported. Beyond adverse event reporting, regulators, globally, are currently supporting a drive towards increasing post-market performance monitoring. There is an open-question as to whether telemetry streaming, which incorporates patient data, will be acceptable under patient rights however.

The current proposal from the USFDA for regulating medical AI takes a lifecycle approach to regulation, this ensures that ongoing software updates will form a core aspect to the certification.(Health 2020) Minor changes to the software are typically allowed without a complete recertification. New features are often classified as major updates and entail a formal certification. The delivery of software updates, under the software lifecycle model, comes with implicit requirements that the updates are distributed to all users. A traditional software product, where each updated version is purchased separately, is only allowed when the absence of updates will not negatively impact patients. The manufacturer is responsible for all old versions of the product until they have been removed from use.

One issue which is specific to medical AI is the topic of shifting clinical standards. Medical diagnostic criteria change over time. Additionally, medical practitioners undergo intensive training early in their careers and frequently drop to a maintenance-level of education as they age. This slows the dissemination of new criteria depending on career phase. A machine learning system can only maintain its performance characteristics on data which is statistically indistinguishable from the data on which it was trained. In technical terms, the data set must be stationary. Unfortunately, medical data is not.

This problem is referred to as distributional or semantic drift.(Challen et al. 2019) In order to mitigate this, prior to launch a plan must be in place, firstly to detect any such changes, and secondly to react safely to them. Detection of semantic drift requires considerable monitoring. In medical data, there is particular risk that un-modelled sources of variance may be particularly difficult to detect. The response plan will be specific to the product. Particular focus must be paid to the timeliness and feasibility of this plan. For example, what will happen to patients in the time period between a shift in diagnostic criteria first occurring, subsequent detection, and finally sufficient new training data being acquired and a product update being released? Medical AI products whose outputs drive the medical process are particularly susceptible to distributional drift. Similar to the 'smart' algorithms, in **Section 2.1.4**, and adaptive AI, in **Section 2.1.5**, this is another example of a coupled feedback system for which particular care must be paid. Causal inference methods are currently the most appropriate approaches for evaluating such systems.(Pearl 2009; Hernan and Robins 2020)

## 2.2. Clinical Validation

In order to validate a medical AI product a clinical trial, or study, is almost always necessary.(Higgins and Madai 2020) The study occurs following the technical engineering process and follows practices much better known among traditional biomedical regulatory agencies. For these guidelines, the focus is on aspects of the clinical trial design which are difficult to translate when separate departments, involving machine learning on one side and the regulatory process on the other, attempt to design a trial together.

Bridging the gap between AI practitioners and a traditional pharma clinical trial design requires the understanding of (i) the target population, and (ii) the intended use. This can be extended to incorporate three further issues, which on the surface appear trivial, but contain hidden dangers. What are appropriate controls for a digital trial? Do the study endpoints match the intended use? And, can a user misunderstand the interface and consequently misuse the product?

The target population in machine learning terms should be understood from a statistical point-of-view. The statistical analyses required to carry out a thorough analysis of in-sample and new data performance should have contributed to the development of an understanding as to whom the technology correctly works upon. This understanding must, however, permeate the clinical trial design process and should be thoroughly investigated by regulators. Limitations on study enrollment, which might make sense for drug development are much harder to justify for medical AI. For example, wealthier subjects will have very different medical histories from poor subjects. It is unethical to limit access to a medical AI device on socio-economic grounds. Therefore, the study population must make up a much broader cross-section of society than that traditionally seen in drug trials.

Voluntary study participants are likely to show better engagement with a digital product than subsequent users, leading to strong questions over what a suitable control should be. Of course, from a safety point-of-view, this point is often deemed to be less relevant if the device in question presents only a low risk to patients. The question of an appropriate control for efficacy depends on the specifics of the market into which the product will be sold. Where no digital solution currently exists regulators frequently accept the present standard-of-care (SoC) as a control. This means the comparison being made is between SoC with and without the addition of the digital product. Products in such a comparison can expect to benefit from an engagement bias, thus overestimating efficacy. This situation is ultimately expected to disappear, particularly as it does not fulfill the clinical equivalence requirements, once more digital products are available as comparable controls.

The intended use is an issue which frequently leads to differences in communication between machine learning experts and regulatory experts. Developing machine learning models frequently leads to multiple levels of abstraction which deviate the actual function of the model from the intended use of the product. This deviation needs to be monitored and assessed. Naturally, in clinical study design, the study endpoints must match the intended use of the product, which is obvious to regulatory experts, but less so to machine learning developers.

Just like the potential for mistakes in communication between machine learning developers and regulatory experts, on model function versus product intended use, there exists a potential for real-world misapplication of the product. Most users do not understand that computer systems, especially intelligent systems such as 'AI', can make mistakes. As a result, users are overly trusting of computers, or conversely, they learn to ignore them. Both of these outcomes should be actively measured in a trial and product design should mitigate them.

## 3. Conclusion

A well-developed medical AI product has the potential to have a huge clinical impact. In order to be allowed onto the market, development must first follow clear regulatory pathways and practices. Following the recent call for Good Machine Learning Practice (GMLP), by the USFDA, this Perspective presents guidelines as to such best practice.

These guidelines follow a bottom-up approach which should be immediately intuitive to machine learning researchers and developers. The focus is on presenting ML development best practices which are compatible with the compilation of a regulatory dossier suitable for regulatory evaluation. This approach provides a necessary bridge between the risk-assessment world of the medical regulators and the technical approaches commonly used in the development and evaluation of trained machine learning models.

Only through the development of better common standards can we bridge the current translation gap, and deliver on the promises of better medical AI for all.

**Table 4:** Glossary of terms.

| | |
|---|---|
| Accuracy | The fraction of predictions for which a model made a correct prediction on a particular data set. |
| Adverse event / outcome | Undesirable medical event or outcome which arises through the use of a medical product. |
| Area under the curve (AUC) | The area under a graph of true positive rates against false positive rates as a discrimination threshold is varied. |
| Algorithm | A set of procedures which may be automated. |
| Artificial intelligence (AI) | A computer system capable of performing specific tasks which usually are thought to require human intelligence. Current usage is largely synonymous with machine learning, however manually constructed expert systems are equally valid forms of AI. |
| Artificial general intelligence (AGI) | Similar to AI but the addressable task set is general rather than specific. |
| ASTM International | Formerly American Society for Testing and Materials. An international standards body. |
| Bayesian method | Bayesian methods are statistical methods under which the conclusions which may be made about a particular parameter or data generating process are made in terms of probability statements which are conditional on the observed values of the dependent measure combined with prior assumptions about the parameter distributions. |
| Confusion matrix | A type of contingency table, for classifier models and diagnostic tools, which summarizes false positives, true positives, false negatives and true negatives for a model applied to a particular data set. |
| CONSORT-AI | Consolidated Standards of Reporting Trials - Artificial Intelligence. |
| Convergence | Both statistical and ML algorithms approach a solution gradually through a process referred to as convergence. Large amounts of data are required to ensure a solution which accurately reflects the underlying properties of the data generating process. |
| Cross-validation | A method for ML model performance evaluation. It involves training a model on a subset of data and testing it on the remaining data. |
| Data generating process | Statistical theory definition of an underlying statistical process which generates observed data. |
| Declaration of conformity (DOC) | A declaration by a manufacturer that the development of a medical product conforms to specific consensus international standards. |
| Design history file (DHF) | A compilation of documentation which describes the design history of a medical device. |
| Device master record | Similar to Technical File. A set of documents that describes a product and can prove that the product was designed and according to the requirements of a quality management system. |
| Efficacy | A medical product must provide a beneficial medical effect. |
| Engagement bias | Patients voluntarily enrolled in studies frequently show higher levels of adherence than a normal population. This greater adherence may lead to demonstrated efficacy which is not reflective of normal product use outside of a trial. Given that digital therapeutics trials often lack proper controls the imbalance is likely to lead to a biased assessment of the efficacy of the therapy over standard-of-care. |
| False negative | A diagnostic which returns a negative classification, e.g. infection not present, when the patient should have been classified as positive. One of the four elements of a confusion matrix. |
| False positive | A diagnostic which returns a positive classification, e.g. infection present, when the patient should have been classified as negative. One of the four elements of a confusion matrix. |
| Generalization | The ability for a ML model to perform with similar metrics, and metric variance, on a test set as it does on the training set. |
| GMLP | Good Machine Learning Practice. A term used by the USFDA. A set of practices which should be followed in the development of products incorporating ML. |
| Harmonised standards | A list of international standards accepted by EU regulators to demonstrate that products comply with legislation. |
| Hyperparameter tuning | The process of tuning the parameters required to control a ML algorithm. |
| IEC | International Electrotechnical Commission. An international standards body. |
| IEEE | Institute of Electrical and Electronic Engineers. An international standards body. |
| Imputation | Mathematical approach to fill-in missing data with inferred values. |

| In-distribution performance | Performance of a model on data sampled using an identical data generating process to that which was used to generate the original data set. In practice, this is a theoretical concept from statistical theory and leads to theoretical-property based and heuristic-based approaches to analysis. |
|---|---|
| In-sample performance | Performance of a model on actual samples from the original data set used in model development. |
| In-vitro Diagnostic (IVD) | Originally used for lab-based in-vitro diagnostics, this regulatory category also encompases digital diagnostics. |
| Input encoding values | Transformation of categorical data to a format compatible with ML algorithm and model inputs. Example, age-ranges may be converted to one-hot encoding. |
| Input plausibility model | An input filter model which examines data inputted to a ML model in order to evaluate whether the model was trained to operate on such data. |
| Intended purpose | The use for which a product is intended according to the manufacturer's description, labelling and sales material. This does not include maintenance, or non user-facing, operations. |
| Intended use | Manufacturer's declaration as to all of the valid usages to which the product may be put. This includes operations such as 'maintenance modes'. The focus is on use, not just purpose. (Intended purpose is a related, but distinct definition). |
| ISO | International organisation for standardization. An international standards body. |
| ITU/WHO | International Telecommunications Union and World Health Organization. An international standards body. |
| Kernel smoother function | A mathematical function which takes multiple inputs, or high-dimensional data, and projects it to a lower dimensional plane. Useful in smoothing input data, however the parameters must be fixed prior to the learning of the weights in the ML model (e.g as part of hyperparameter tuning). |
| Locked ML-model | The term 'locked' is a medical device usage meaning that the model does not change once it is certified for deployment. In practice this means that a version specific ML-model is trained and the product validation and certification is tied to the use of this particular model version. |
| Machine learning (ML) | The study of computer algorithms that improve automatically through experience and by the use of data. |
| Machine learning - (trained) models | Trained models take data similar to that in the training set and make predictions based on those similarities. |

| Machine learning - algorithms | Computer algorithms for learning from data. |
| --- | --- |
| Machine learning - products | Products which incorporate machine learning models or algorithms in their operation. |
| Medical AI products | Products which use artificial intelligence on bio-medical data. |
| Medical Device (MD) | Under EU law, a medical device is a legal category, which must be certified before being sold. |
| MINIMAR | MINimum Information for Medical AI Reporting. |
| Model (e.g. ML / mathematical model) | A mathematical equation, or set of equations, which numerically transforms inputs into outputs. |
| Model Cards | Short documents accompanying trained machine learning models that provide benchmarked evaluation in a variety of conditions that are relevant to the intended application domain. |
| Multi-fold cross-validation | The cross-validation approach is repeated multiple times with independent samples for the training and test sets each time. |
| Negative predictive value (NPV) | Given that a negative classification has been made, e.g. by a ML model, this is the likelihood of it being a true negative. Dependent on prevalence. |
| NEMA | National Electrical Manufacturers Association. An international standards body. |
| Non-stationarity of data | Data whose long-term distribution changes over time. (It drifts). |
| Numerical sensitivity analysis | A numerical approach to studying how uncertainty about differences in the outputs of a model may be attributed to different sources of uncertainty in the inputs. |
| One-hot encoding | An input encoding technique which transforms a single variable, which supports multiple values, into multiple variables only one of which takes on the value 'true' for a given valid input. |
| Online learning algorithm | One form of algorithm which may be used to deploy adaptive AI solutions. Every new data point leads to a model update. These algorithms are typically used in financial applications to track time series. |
| Performance metrics | Numerical evaluations of the ability of a model to carry out its task, on a given data set, according to certain criteria. |
| Positive predictive value (PPV) | Given that a positive classification has been made, e.g. by a ML model, this is the likelihood of it being a true positive. Dependent on prevalence. |
| Premarket approval (PMA) | The most stringent form of device marketing authorization required by the USFDA. For high-risk medical devices the |

| | manufacturer must apply for permission prior to carrying out any marketing activity. |
|---|---|
| Premarket authorization filing | A set of documents which must be submitted to the USFDA in order to obtain premarket approval (PMA). |
| Probably approximately correct (PAC) framework | A framework for mathematical analysis of machine learning algorithms. |
| QA | Quality assurance. |
| QMS | Quality management system. |
| Recognized consensus standards | A list of national and international standards, accepted by the USFDA, for which a Declaration of Conformity may be made as part of the premarket authorization filing. |
| Regulatory dossier | The documentation supporting a regulatory application. |
| Safety | A product must be safe for users and other stakeholders. It should not cause unnecessary harm. This does not preclude the development of products which provide more benefit than harm. |
| Semantic drift | The meaning of medical terms changes over time. Often this is too slow for human perception but is highly detrimental for ML models. |
| Sensitivity | A phrase used in diagnostics. The proportion of positive cases which are correctly identified. |
| Shapley values | A mathematical approach derived using game theory for determining the key factors driving decisions. Growing use in evaluation of ML models. |
| Software as a Medical Device (SaMD) | Defined by the international medical device regulators forum as, "software intended to be used for one or more medical purposes that perform these purposes without being part of a hardware medical device."(IMDRF SaMD Working Group 2013) |
| Specificity | A phrase used in diagnostics. The proportion of negative cases which are correctly identified. |
| SPIRIT-AI | Standard Protocol Items: Recommendations for Interventional Trials - Artificial Intelligence |
| Stakeholder requirements | A list of requirements in the form of statements on behalf of product stakeholders. |
| Standard of care (SoC) | The standard treatment path for a given clinical indication, typically laid down by an expert body either for a large hospital or on a national basis. |

| Standard operating procedure (SOP) | For every circumstance encountered in either the development or operation of a product there should be a written operating procedure. |
|---|---|
| Support vector machine | A ML algorithm used to create ML models from data. |
| Surveillance | A regulatory affairs term which refers to the monitoring of the real-world performance of a medical device, IVD or drug with particular focus on detecting adverse events. |
| Synthetic data | Data which has been artificially generated to strongly resemble the data on which a ML model was trained. |
| Teaching signal | The feedback signal which is an input to a reward learning algorithm. Typically this begins as a yes/no feedback as to whether existing performance on specific examples was correct or not. The algorithm then uses this signal to adapt the ML model. |
| Technical file | A set of documents that describes a product and can prove that the product was designed according to the requirements of a quality management system (QMS). |
| Technical requirements | An exhaustive list of requirements, as to what a product should and should not do. |
| Training set | A set of data points used as input to a ML algorithm in order to produce a trained ML model as an output. |
| Tree node | A single decision point in a tree-based decision model. Tree models arrive at an output by sequentially performing the decisions represented by the tree. |
| TRIPOD | Transparent Reporting of a multivariable prediction model for Individual Prognosis Or Diagnosis |
| True negative | A diagnostic which returns a negative classification, e.g. infection not present, when the patient should have been classified as negative. One of the four elements of a confusion matrix. |
| True positive | A diagnostic which returns a positive classification, e.g. infection present, when the patient should have been classified as positive. One of the four elements of a confusion matrix. |
| Unlocked ML-model | The USFDA defines this as a ML-model which is coupled to an ongoing learning process via a ML-algorithm. This means that the model may not give the same set of results for identical inputs presented at different points in time. |
| Unsupervised algorithm | A ML algorithm which works on unlabelled data. That is data for which the target model output is currently unknown. |

| User Interface / User Experience (UI/UX) | Terms commonly used in product design to express the user perspective of an interface or product experience. |
|---|---|
| USFDA | US Food and Drug Administration. Responsible for regulating medical AI in the USA. |
| Validation | The assurance that a product meets the needs of the customer and other identified stakeholders. |
| Verification | The evaluation of whether or not a product complies with a requirement, specification, or imposed condition. |

**Acknowledgements**

The author wishes to thank Prof. Christian Johner, Dr. Florian Aspart, Dr. Richard Tomsett, Richard Baxter, Oliver Rieger and Jonas Seiler for their help in reviewing early drafts of this manuscript. Also, many thanks to the anonymous reviewers who gave extremely helpful and detailed feedback on the submitted draft.